\documentclass[10pt,letterpaper,twocolumn]{revtex4}

\usepackage{amsmath}
\usepackage{graphicx}
\usepackage{amssymb}

\begin{document}

%\twocolumn[ %% activate for two-column option

%%%%%%%%%%%%%%%%%% title page information %%%%%%%%%%%%%%%%%%
\title{Non-Gaussian Statistics of Multiple Filamentation}

\author{Pavel M. Lushnikov and Natalia Vladimirova}

\address{Department of Mathematics and Statistics,
University of New Mexico, Albuquerque, NM 87131, USA}

%\email{plushnik@math.unm.edu} %% email address is required

 %\homepage{http://math.unm.edu/\textasciitilde% $\sim$ %\~{ } %\textasciitilde
 %plushnik/} %% author's URL, if desired

%%%%%%%%%%%%%%%%%%% abstract and OCIS codes %%%%%%%%%%%%%%%%
%% [use \begin{abstract*}...\end{abstract*} if exempt from copyright]

\begin{abstract}
We consider the statistics of light amplitude fluctuations for the
propagation of a laser beam subjected to multiple filamentation in an
amplified Kerr media, with both linear and nonlinear dissipation.
Dissipation arrests the catastrophic collapse of filaments, causing
their disintegration into almost linear waves.  These waves form a
nearly-Gaussian random field which seeds new filaments. For small
amplitudes the probability density function (PDF) of light amplitude is
close to Gaussian, while for large amplitudes the PDF has a long
power-like tail which corresponds to strong non-Gaussian fluctuations,
i.e. intermittency of strong optical turbulence.  This tail is determined by the universal form of near singular filaments and the
PDF for the maximum amplitudes of the filaments.

{\small OCIS codes: {(190.0190)   Nonlinear optics; (190.5940);   Self-action effects;  (260.5950)   Self-focusing}}
\end{abstract}

%\ocis{(190.0190)   Nonlinear optics; (190.5940);   Self-action effects;  (260.5950)   Self-focusing}
%\pacs{42.65.Sf,  42.65.Jx}
%42.65.Sf 	Dynamics of nonlinear optical systems; optical instabilities, optical chaos and complexity, and optical spatio-temporal dynamics
%42.65.Jx 	Beam trapping, self-focusing and defocusing; self-phase modulation
%47.27.-i 	Turbulent flows
%52.38.Hb 	Self-focussing, channeling, and filamentation in plasmas

%%%%%%%%%%%%%%%%%%%%%%% References %%%%%%%%%%%%%%%%%%%%%%%%%

\maketitle

%]

The self-focusing and multiple filamentation of an intense laser beam
propagating through a Kerr media has been the subject of intense research
since the advent of lasers \cite{Boyd2003,SulemSulem1999}.  Multiple
filamentation has many applications ranging from laser fusion
\cite{LindlPhysPlams2004,LushnikovRosePRL2004,LushnikovRosePlasmPysContrFusion2006}
to the propagation of ultrashort pulses in the atmosphere
\cite{MlejnekKolesikMoloneyWrightPRL1999,BergeEtAlRepProgrPhys2007}.
Here we study the statistics of multiple filamentation, which can be
viewed as an example of strong optical turbulence with intermittency
\cite{DyachenkoNewellPushkarevZakharovPhysicaD1992},
i.e. strong non-Gaussian fluctuations of the amplitude of the laser field
\cite{FrischBook1995}.

%Equations do not have tails; a tail is a feature of the PDF.  The PDF
%describes the statistics of something? What is the something?

Long non-Gaussian tails of the PDF of light amplitude fluctuations have
been previously observed in filamentation experiments
\cite{MontinaBortolozzoResidoriArecchiPRL2009} and optical rogue waves
\cite{SolliRopersKoonathJalaliNature2007}.  Long tails were obtained
in solutions of the complex Ginzburg-Landau equation with a quintic nonlinearity
\cite{92IT}.  The analytical form of a long tail of PDF for velocity gradient
 dominated by near-singular
shocks was obtained in solutions of the forced Burgers equation \cite{97EKMS}.

Here we describe the propagation of a laser beam
through the amplified Kerr media by the regularized nonlinear
Schr\"odinger  equation (RNLS) in dimensionless form,
\begin{eqnarray}\label{nlsregul}
  i\partial_z\psi+(1-ia\epsilon)\nabla^2\psi+(1+ic\epsilon )|\psi|^{2}\psi=
  i\epsilon b \psi,
\end{eqnarray}
where the beam is directed along $z$-axis, ${\bf r}\equiv (x,y)$ are the
transverse coordinates, $\psi({\bf r},z)$ is the envelope of the
electric field, and $\nabla\equiv\left ( \frac{\partial}{\partial x},
\frac{\partial}{\partial y}\right )$.  The term with $a$ describes a
wavenumber-dependent linear absorption, the term with $c$ corresponds
to two-photon absorption, and $b$ is the linear gain coefficient.  It
is assumed that the wavenumber-independent part of linear absorption
is included in $b$ so that the $a$-term corresponds to the expansion
of the general absorption coefficient near the carrier wavenumber of
the laser beam in the Fourier domain.  Here, $\epsilon\ll 1$ and we
generally assume $a, \, c, \, b \sim 1.$ RNLS
(\ref{nlsregul}) is also called the complex Ginzburg-Landau equation.

RNLS  (\ref{nlsregul}) can be realized experimentally in
numerous systems, including e.g. the propagation of a laser beam in a
ring cavity with a thin slab of Kerr media and amplification.  In this
case the nonlinear phase shift of the laser beam at each round trip is
small so we can obtain (\ref{nlsregul}) in a mean-field approximation
\cite{TlidiHaeltermanMandelEurophysLett1998,LushnikovSaffmanPRE2000}
with $z$ corresponding to the number of round trips in the cavity.
RNLS  (\ref{nlsregul}) also describes the multiple
filamentation of an intense, ultrashort laser beam in a Kerr media if
we average over the temporal extent of the pulse
\cite{TzortzakisBergeCouaironFrancoPradeMysyrowiczPRL2001}.
E.g., multiple filamentation experiment  \cite{GuyonBergeEtAlPRA2006} with two-photon-dominated absorption
corresponds to $c\epsilon\simeq 0.025$.

Neglecting dissipation and amplification, we recover the nonlinear
Schr\"odinger equation (NLS)
\begin{equation}\label{nlssigma1}
 i\partial_z\psi+\nabla^2\psi+|\psi|^{2}\psi=0.
\end{equation}
NLS describes a catastrophic collapse (also called wave collapse) of filaments, $\max
\limits_{\bf r} |\psi|\equiv |\psi|_{max}\to \infty$, in a finite
distance along $z$, if the optical power $N=\int |\psi|^2 d{\bf r}$ is
above the critical power $N_c\simeq 11.701$
\cite{SulemSulem1999}. %VlasovPetrishchevTalanovRdiofiz1971}.

The optical power is not conserved in  RNLS
(\ref{nlsregul}) for $\epsilon\ne 0$.  If $b>0$, the amplification
term on the right hand side of (\ref{nlsregul}) results in an
increase of $N$.  If $b=0$ we assume that $N\gg N_c$ for $z=0$ (e.g. $N/N_c\sim 10^4$ in \cite{GuyonBergeEtAlPRA2006}).  In
both cases the modulational instability~\cite{SulemSulem1999} leads to
the growth of
%\cite{BespalovTalanovJETPLett1966} results in growth of transverse
perturbations of the beam and seeds multiple collapsing filaments.  These two cases are called forced and decaying turbulence,
respectively, in reference to turbulence in the Navier-Stokes
equations \cite{FrischBook1995}.  The statistical properties of these
two cases are similar, provided in the decaying case we consider
distances along $z$ at which a relative cumulative decay of $N$ is
small.

\begin{figure}%[htbp]
\centering\includegraphics[width =8.3cm]{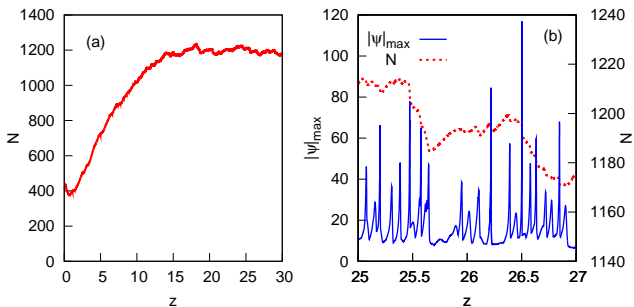}
\caption{(Color online) (a)
$N(z)$ from simulation of RNLS (\ref{nlsregul})
with $b=20$,  $a=c=1$,  $\epsilon=0.01$.
(b) The zoom of $N(z)$ in a smaller interval in $z$ (solid curve,
scale on the right) superimposed the $ |\psi|_{max}(z)$ (dashed
curve, scale on the left).  All simulations used the fourth-order pseudo-spectral split-step algorithm
%\cite{Yoshida1990}.
on %square domain
$-12.8\le x,y\le 12.8$ with periodic boundary conditions
at resolution $4096\times 4096$ grid points.
Initial conditions were a superposition of 100 randomly placed real-valued Gaussians with amplitudes and radii on $[-2,2]$  and
$[1,2]$.
}
\label{fig:fig1N}
\end{figure}

\begin{figure}%[htbp]
\centering\includegraphics[width =6.3cm]{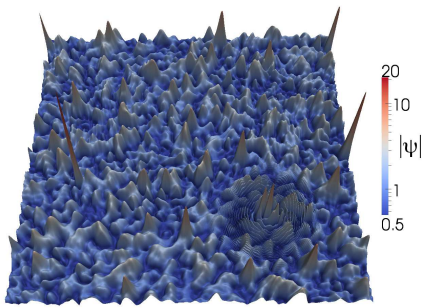}
\caption{(Color online) Snapshot of $|\psi|$ (vertical axis) vs. spatial coordinates $(x,y)$ for simulation of Figure \ref{fig:fig1N}.} \label{fig:figpsi}
\end{figure}

In this Letter we focus on a forced case in which a dynamic balance is
achieved between the pumping of optical power into the laser beam and
dissipation.  Figure \ref{fig:fig1N}(a) shows the
evolution of $N(z)$ obtained from a numerical solution of  RNLS
 (\ref{nlsregul}).  The optical power grows until reaching a
statistical steady-state corresponding to fully developed optical
turbulence with $N\simeq 1200$.  In this regime the amplitude $|\psi|$
is characterized by the random distribution of filaments in ${\bf r}$
and $z$, as seen in the snapshot of $|\psi|$ for a fixed $z$ in Figure
\ref{fig:figpsi}.  Dissipation is important only
when the amplitude of each collapsing filament is near to its maximum (see Figure
\ref{fig:fig1N}(b)) and as well as for large wavenumbers $k$.  When
$|\psi|_{max}(z)$ goes through a maximum, $N$ experiences a fast decay
due to dissipation.  The influence of periodic boundary conditions on
the statistical properties of optical turbulence can be neglected
if the simulation domain is large enough, so that $N\gg N_c$ (in Figure
\ref{fig:fig1N}, $N\sim 100 N_c$).

The evolution of each collapsing filament is well approximated for large $|\psi|$ by a
self-similar radially-symmetric solution
\cite{SulemSulem1999,FibichPapanicolaouSIAMJApplMAth1999}:
\begin{equation}\label{selfsimilargamma}
\begin{split}
|\psi({\bf r},z)|=\frac{1}{L(z)}R_0(\rho), \quad
\rho=\frac{r}{L(z)}, \quad r\equiv |{\bf r}|,
\end{split}
\end{equation}
where $L(z)$ is the transverse spatial scale of a
filament and $R_0$ is the ground state soliton solution of
 NLS (\ref{nlssigma1}), given by
%
%\begin{eqnarray}\label{R0eq}
$\nabla^2R_0-R_0+R_0^{3}=0,$
%\end{eqnarray}
%
and corresponding to the critical power, $N_c=\int R_0(r)^2d{\bf r}$
\cite{SulemSulem1999}.

\begin{figure}%[htbp]
\centering\includegraphics[width =8.3cm]{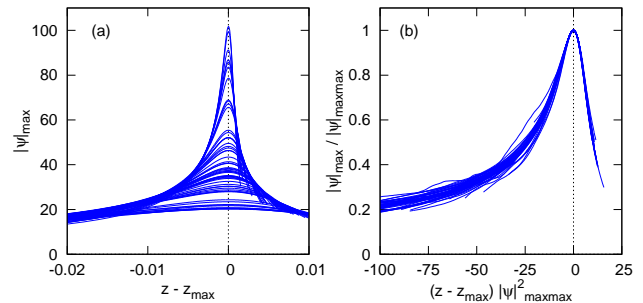}
\caption{(Color online)
Dependence of $|\psi|_{max}(z)$ for multiple individual collapsing filaments (a)
in the non-rescaled units; (b) the rescaled units (see the text for the
description of rescaling).  Individual filaments are extracted from
simulation of RNLS  (\ref{nlsregul}) with parameters of Figure
\ref{fig:fig1N}.} \label{fig:psigamma}
\end{figure}

If $\epsilon=0$ then $L(z)\simeq (2\pi)^{1/2}
({z_0-z})^{1/2}/(\ln|\ln(z_0-z)|)^{1/2}$
describes a singularity (catastrophic collapse of a filament) as $z \rightarrow z_0$
\cite{SulemSulem1999,FibichPapanicolaouSIAMJApplMAth1999}.
%\cite{FraimanJETP1985,LandmanPapanicolaouSulemSulemPRA1988,BLeMesurierPapanicolaouSulemSulemPhysD1988},
For $\epsilon\neq 0$ the collapse is regularized and
$|\psi|_{max}(z)$ achieves a maximum $|\psi|_{maxmax}$ at some
$z=z_{max}$.  A function
\begin{equation}\label{gammadef}
\gamma\equiv L\frac{dL}{dz},
\end{equation}
changes slowly with $z$ compared to $L$ at $z\lesssim z_{max}$.  In
the vicinity of a collapse, the forcing term in the right hand side of
RNLS  (\ref{nlsregul}) can be neglected; the resulting
equation can be written in rescaled units $ z|\psi|_{maxmax}^2,$ $
{\bf r}|\psi|_{maxmax}$, and $ \psi/|\psi|_{maxmax}$. (Here, we have
also shifted $z_{max}$ to $z=0$.)  As shown in Figure~\ref{fig:psigamma},
$|\psi|_{max}(z)$ rescaled in these units exhibits a universal behavior
for all near-singular filaments, --- even for $\epsilon\neq0$, and
independent of the complicated structure of optical turbulence.  This
universality is a characteristic feature of two-photon absorption term
in  RNLS  (\ref{nlsregul}), but may not hold for other
types of absorption.

\begin{figure}%[htbp]
\centering\includegraphics[width =8.3cm]{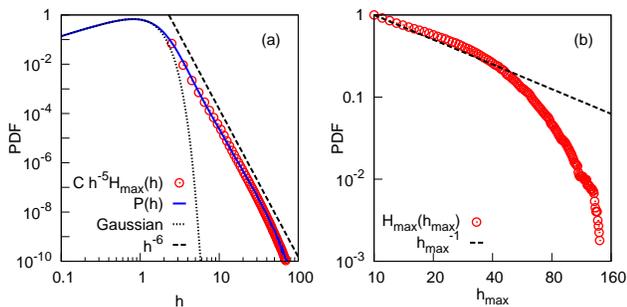}
\caption{(Color online)
   (a) ${\cal P}(h)$ for $|\psi|=h$ (solid curve) for the same
   simulation as in Figure \ref{fig:fig1N}. Dashed line shows fit to
   the Gaussian distribution and dotted line shows $h^{-6}$ power law.
   Circles correspond to the solution of~(\ref{hpower}).  (b)
   $H_{max}(h_{max})$ for $|\psi|_{maxmax}$.  2231 collapse events with
   $|\psi|_{maxmax}>10$ are included in simulations in log-log scale
   (red dots). Dotted line shows $h^{-1}$ power law.  Scattering of
   data points for $h\gtrsim 110$ is due to lack of statistical
   ensemble for large collapses and is reduced for larger simulation times. } \label{fig:PDFsmallpsi}
\end{figure}

Once the amplitude of a filament reaches its maximum, the
 amplitude decreases and subsequently the filament decays into outgoing cylindrical waves as seen in Figure
\ref{fig:figpsi}.  Superposition of these almost linear waves forms a
nearly random Gaussian field and seeds new filaments.  Figure
\ref{fig:PDFsmallpsi} shows the probability ${\cal P}(h)$ for the
amplitude $|\psi|$ to have a value $h$,
% matched with the Gaussian distribution ${\cal P}(h)$ is
determined from simulations as
\begin{equation}\label{PDFabspsidef}
{\cal P}(h)=\frac{\int \delta (|\psi({\bf r},z)|-h)d{\bf r} dz}
{\int d{\bf r}dz}.
\end{equation}
Here, the integrals are taken over all values of ${\bf r}$ and all
values of $z$ after the turbulence has reached the statistically
steady state.  We observe that the fit to the Gaussian distribution
works very well for $|\psi|\lesssim 2$ which corresponds to almost
linear waves, while for $|\psi|\gtrsim 3$ the PDF has a power law-like
dependence indicating  intermittency
\cite{ChungLushnikovVladimirovaAIPProc2009}.

We now show that the power-like tail of ${\cal P}(h)$ results from the
near-singular filaments.  This approach dates back to the idea of
describing strong turbulence in the Navier-Stokes equations through
singularities of the Euler equations \cite{FrischBook1995}.
Unfortunately, this hydrodynamic problem remains unsolved.  The forced
Burgers equation remains the only example of an analytical description
of strong turbulence in which the tail of the PDF for negative
gradients follows a well established $(-7/2)$ power law \cite{97EKMS},
dominated by the dynamics of near-singular shocks.

%There are two major contribution to PDF  ${\cal P}$.

First we calculate the contribution to the PDF from individual collapsing filaments.
As shown in Figure \ref{fig:psigamma}, the filament
amplitude $|\psi|_{max}$ reaches the maximum $|\psi|_{maxmax}\equiv
h_{max}$ at $z=z_{max}$, and rapidly decays for $z>z_{max}$. While
neglecting the contribution to ${\cal P}(h)$ from $z\gtrsim z_{max}$,
we calculate the contribution of an individual filament to ${\cal P}(h)$
through the conditional probability ${\cal P}\left(h|h_{max}\right)$
using (\ref{selfsimilargamma}), (\ref{PDFabspsidef}), and (\ref{gammadef})
as follows
 \begin{eqnarray}
 \begin{split}
&{\cal P}\left(h|h_{max}\right)\propto \int\limits
^{z_{max}} dz \int d{\bf
r}\delta\left(h-\frac{1}{L(z)}R_0\left(\frac{r}{L(z)}\right)\right) \\
&\propto
\int d{\bf
\rho}\, \rho\int\limits^{L(z_{max})} \frac{dL \,
L^{3}}{\gamma}\delta\left(h-\frac{1}{L(z)}R_0(\rho)\right) \\
&\simeq\int
\frac{d{\bf\rho}\,\rho}{{ \langle \gamma \rangle}h^{5}}
%\times
{ [R_0(\rho)]^{4}}
\Theta\left(\frac{R_0(0)}{L(z_{max})}-h\right)
 %\nonumber \\
%&=& Const \ \Theta\left(\frac{R_0(0)}{L(z_{max})}-h\right)
%\frac{1}{h^{5}}.
\\
&= Const \  h^{-5} \ \Theta\left(h_{max}-h\right),
\label{P_L_rho}
\end{split}
\end{eqnarray}
where $h_{max}=R_0(0)/L(z_{max})$, and $\Theta(x)$ is the Heaviside
step function.  Here, we have changed the integration variable from $z$ to
$L$ and approximated $\gamma(z)$ under the integral by its average
value $\langle \gamma \rangle$ as $\gamma(z)\simeq \langle \gamma
\rangle\sim-0.5$.  This approximation is valid for $z\lesssim
z_{max}$ outside the neighborhood of $z=z_{max}$.

As a second step we calculate ${\cal P}(h)$ by integration over all
values of $h_{max}$ using equation (\ref{P_L_rho}) as follows
\begin{eqnarray}\label{hpower}
{\cal P}(h)&=&\int  dh_{max}{\cal P}(h|h_{max}) {\cal P}_{max}(h_{max})
\nonumber
\\
&\simeq&  Const \ h^{-5}\int dh_{max} \Theta(h_{max}-h) {\cal P}_{max}(h_{max})\nonumber \\
&=& Const \ h^{-5} H_{max}(h),
\end{eqnarray}
where ${\cal P}_{max}(h_{max}) $ is the PDF for $h_{max}=|\psi|_{maxmax}$
and $H_{max}(h)\equiv \int^\infty_{h} {\cal P}_{max}(h_{max})
dh_{max}$ is the cumulative probability that $|\psi|_{maxmax}>h$.

Figure~\ref{fig:PDFsmallpsi}(b) shows $H_{max}(h_{max})$. Circles in
Figure~\ref{fig:PDFsmallpsi}(a) show the prediction of
equation~(\ref{hpower}) with $H_{max}(h_{max})$ from Figure
\ref{fig:PDFsmallpsi}(b).  The constant in equation~(\ref{hpower})
%(\ref{fig:fig1N})
was chosen to fit the circles and the solid curve in Figure
\ref{fig:PDFsmallpsi}(a).  The very good agreement between
these two curves, first, justifies the assumptions used in derivation
of the equation~(\ref{hpower}),
%(\ref{fig:fig1N})
and second, shows
that the intermittency of optical turbulence of  RNLS
(\ref{nlsregul}) is due to collapse dynamics, which is the main result
of this Letter. Figure \ref{fig:PDFsmallpsi} also shows that
$H_{max}(h_{max})$ is not well approximated by $h_{max}^{-1}$ (and is
not universal because it depends on the parameters $a, \,b, \, c$ and
$\epsilon$.)  Consequently, $h^{-6}$ is only a crude approximation for
${\cal P}(h)$.

%7)Now we show that PDF ... is dominated by two effects. First is individual collapse and second is the contribution from PDF of collapse maximum.

%\section{Conclusion}

\section*{Acknowledgments}

Support was provided by NSF grant DMS
 0807131 and UNM RAC grant.

%%%%%%%%%%%%%%%%%%%%%%%%%%  body  %%%%%%%%%%%%%%%%%%%%%%%%%%

\end{document}